\def\etal{{et al.\thinspace}}
\def\mearth{{\rm\,M_\oplus}}

\def\bbc{{\beta/\beta_{crit}}}

\documentclass[apjl]{emulateapj}

\received{}
\accepted{}

\lefthead{}
\righthead{}

\begin{document}

\shorttitle{}
\shortauthors{Raymond et al.}

\title{Planet-planet scattering leads to tightly packed planetary systems}

\author{Sean N. Raymond\altaffilmark{1,2}, 
Rory Barnes\altaffilmark{2,3},
Dimitri Veras\altaffilmark{4},
Philip J. Armitage\altaffilmark{5},
Noel Gorelick\altaffilmark{6}
\& Richard Greenberg\altaffilmark{7}}

\altaffiltext{1}{Center for Astrophysics and Space Astronomy, 389 UCB,
University of Colorado, Boulder CO 80309; sean.raymond@colorado.edu}
\altaffiltext{2}{Virtual Planetary Laboratory}
\altaffiltext{3}{Department of Astronomy, University of Washington, Seattle, WA 98195}
\altaffiltext{4}{Astronomy Department, University of Florida, Gainesville, FL 32111}
\altaffiltext{5}{JILA, University of Colorado, Boulder CO 80309}
\altaffiltext{6}{Google, Inc., 1600 Amphitheatre Parkway, Mountain View, CA 94043}
\altaffiltext{7}{Lunar and Planetary Laboratory, University of Arizona, Tucson, AZ }

\begin{abstract}
The known extrasolar multiple-planet systems share a surprising dynamical
attribute: they cluster just beyond the Hill stability boundary.  Here we show
that the planet-planet scattering model, which naturally explains the observed
exoplanet eccentricity distribution, can reproduce the observed distribution
of dynamical configurations.  We calculated how each of our scattered systems
would appear over an appropriate range of viewing geometries; as Hill
stability is weakly dependent on the masses, the mass-inclination degeneracy
does not significantly affect our results.  We consider a wide range of
initial planetary mass distributions and find that some are poor fits to the
observed systems.  In fact, many of our scattering experiments overproduce
systems very close to the stability boundary.  The distribution of dynamical
configurations of two-planet systems actually may provide better
discrimination between scattering models than the distribution of
eccentricity.  Our results imply that, at least in their inner regions which
are weakly affected by gas or planetesimal disks, planetary systems should be
``packed'', with no large gaps between planets.
\end{abstract}


\section{Introduction}

The observed eccentricities of extra-solar planets can be readily explained by
a simple model that assumes that virtually all planetary systems undergo
dynamical instabilities (Ford \etal 2003; Adams \& Laughlin 2003; Chatterjee
\etal 2008; Juric \& Tremaine 2008; Ford \& Rasio 2008).\footnote{Several
other models to explain the extra-solar eccentricity distribution exist; see
Ford \& Rasio (2008) for a summary.}  In the context of this model, planetary
systems are expected to form in marginally stable configurations, meaning that
they are stable for at least the timescale of rapid gas accretion of $\sim
10^5$ years (Pollack \etal 1996) but ultimately unstable, probably on a
timescale comparable to the gaseous disk's lifetime of $\sim 10^6$ years (Haisch \etal
2001).  This instability timescale implies an initial separation between
planets of perhaps 4-5 mutual Hill radii $R_{H,M}$, where $R_{H,M} = 0.5 \,
(a_1+a_2)[(M_1+M_2)/3M_\star]^{1/3}$; $a_1$ and $a_2$ are the orbital
distances, $M_1$ and $M_2$ are the masses of two adjacent planets, and
$M_\star$ is the stellar mass (Chambers \etal 1996; Marzari \& Weidenschilling
2002; Chatterjee \etal 2008).\footnote{For Jupiter-mass planets, separations
of $\sim 4-5 R_{H,M}$ are close to the 3:2 and 2:1 mean motion
resonances. Thus, an alternate argument in favor of planets forming with such
spacings invokes resonant capture (Snellgrove \etal 2001) followed by
turbulent removal from resonance (Adams \etal 2008) during the gaseous disk
phase.}  After a delay of $10^5-10^6$ years, a typical system of three or more
planets with separations of $4-5 R_{H,M}$ becomes unstable, leading to close
encounters between two planets, strong dynamical scattering, and eventual
destruction of one or two planets by either collision with another planet,
collision with the star, or, most probably, hyperbolic ejection from the
system (Rasio \& Ford 1996; Weidenschilling \& Marzari 1996; Lin \& Ida 1997;
Papaloizou \& Terquem 2001).  It is the planets that {\em survive} the
dynamical instability that provide a match to the observed extra-solar
eccentricities.

Additional dynamical information can be obtained from the known extra-solar
multiple planet systems.  In particular, the stability in two-planet systems
can be guaranteed for planets with particular masses and orbital
configurations.  The edge of stability can be quantified in terms of the
proximity to the analytically-derived Hill stability limit using the
dimensionless quantity $\bbc$ (the stability boundary is located at $\bbc =
1$; see Section 3).  Dynamical analyses have shown that the known
multiple-planet systems are clustered just beyond the edge of stability (i.e.,
at $\bbc \gtrsim 1$; Barnes \& Quinn 2004; Barnes \& Greenberg 2006, 2007).

In this paper we study the stability of the surviving planets in several
thousand 3-planet systems that have undergone planet-planet scattering leading
to the loss of one planet.  We find that in the aftermath of dynamical
instabilities, the surviving planets cluster just beyond the stability
boundary, providing a good match to the observed
values.  This provides support for planet-planet scattering as an active
process in extra-solar planetary systems.  This result also has
consequences for the packing of planetary systems and the ``Packed Planetary
Systems'' hypothesis (Barnes \& Raymond 2004; Raymond \& Barnes 2005; Raymond
\etal 2006; Barnes \etal 2008).  The paper proceeds as follows: we describe
our scattering simulations (\S 2), summarize Hill stability theory and define
$\bbc$ (\S 3), present our results (\S 4) and discuss the consequences (\S 5).

\section{Scattering Simulations}

Our scattering simulations are drawn from the same sample as in Raymond \etal
(2008a).  Each simulation started with three planets randomly separated by 4-5
mutual Hill radii.  The three planets were placed such that the outermost
planet was located two (linear) Hill radii $R_H$ interior to 10 AU ($R_H = a
[M/3 M_\star]^{1/3}$).  We performed ten sets of simulations, varying the
planetary mass distribution.  For our two largest sets (1000 simulations each)
we randomly selected planet masses according to the observed distribution of
exoplanet masses: $dN/dM \propto M^{-1.1}$ (Butler \etal 2006).  In the {\tt
Mixed1} set we restricted the planet mass $M_p$ to be between a Saturn mass
$M_{Sat}$ and three Jupiter masses $M_{Jup}$.  For our {\tt Mixed2} set, the
minimum planet mass was decreased to 10 $\mearth$.  We also performed four
{\tt Meq} sets (500 simulations each) with equal mass planets for $M_p = 30
\mearth$, $M_{Sat}$, $M_{Jup}$, and $3 M_{Jup}$. Finally, the {\tt Mgrad} sets
(250 simulations each) contained radial gradients in $M_p$.  For the JSN set,
in order of increasing orbital distance, $M_p$ = $M_{Jup}$, $M_{Sat}$, and $30
\mearth$.  For the NSJ set, these masses were reversed, i.e., the $M_{Jup}$
planet was the most distant.  The 3JJS and SJ3J sets had, in increasing radial
distance, $M_p$ = $3 M_{Jup}$, $M_{Jup}$ and $M_{Sat}$, and $M_p$ = $M_{Sat}$,
$M_{Jup}$ and $3 M_{Jup}$, respectively.

Planetary orbits were given zero eccentricity and mutual inclinations of less
than 1 degree.  Each simulation was integrated for 100 Myr with the hybrid
{\em Mercury} integrator (Chambers 1999) using a 20 day timestep.  We required
that all simulations conserve energy to better than ${\rm d}E/E < 10^{-4}$,
which is needed to accurately test for stability (Barnes \& Quinn 2004).  We
achieved this by reducing the timestep to 5 days for simulations with ${\rm
d}E/E > 10^{-4}$ and then removing simulations that still conserved energy
poorly.  As expected, these systems were typically unstable on $10^5-10^6$
year timescales.  In addition, about 1/4 of simulations were stable for 100
Myr which shows that we started close to the stability boundary.  For this
paper, we restrict our analysis to the subsample of simulations that 1) were
unstable, and 2) contained two planets on stable orbits after 100 Myr (i.e.,
one and only one planet was destroyed).

\section{Hill Stability}

For the case of two planets with masses $M_1$ and $M_2$ orbiting a star,
dynamical stability is guaranteed if:
\begin{eqnarray}
\frac{-2 (M_\star+M_1+M_2)}{G^2 (M_1 M_2 + M_\star M_1 + M_\star M_2)^3} \,
c^2 h \geq 
\nonumber \\
1+3^{4/3} \frac{M_1 M_2}{M_\star^{2/3} (M_1+M_2)^{4/3}} -
\frac{M_1 M_2 (11 M_1 + 7 M_2)}{3 M_\star (M_1 +M_2)^2}, 
\end{eqnarray}
\noindent where $c$ and $h$ represent the total orbital angular momentum and
energy of the system, respectively (Marchal \& Bozis 1982; Gladman 1993; Veras
\& Armitage 2004; note that this definition assumes that $M_1 > M_2$).  We
refer to the left side of Eqn 1 as $\beta$ and the right side as
$\beta_{crit}$ (Barnes \& Greenberg 2006).  The quantity $\bbc$ therefore
measures the proximity of a pair of orbits to the Hill stability limit of
$\bbc=1$.  We note that our $\bbc$ analysis only applies for two-planet
non-resonant systems, because perturbations from additional companions can
shift the stability boundary to values other than 1 (Barnes \& Greenberg 2007).

When calculating $\bbc$ for extra-solar systems, past research (Barnes
\& Greenberg 2006, 2007) has assumed coplanar orbits with masses equal
to minimum masses.  Those values of $\bbc$ were systematically affected by the
mass-inclination degeneracy, probably resulting in overestimations. In
contrast, our simulations provide the full three-dimensional orbits, and hence
we can calculate the true value of $\bbc$. More importantly, if we assume that
viewing geometries are distributed isotropically (i.e. edge-on systems are
more likely than face-on), we can determine how $\bbc$ would be calculated
from radial velocity data (e.g. assuming coplanar, edge-on orbits). For
example, if two planets with masses $M_b$ and $M_c$ have inclinations
(relative to their invariable plane) $i_b$ and $i_c$, and the inclination to
the line of sight is $I$, then the ``observed'' $\bbc$ would use masses
$M_b\sin (i_b + I)$ and $M_c\sin (i_c + I)$. In \S 4 we use this approach to
build a distribution of $\bbc$ that is directly comparable to the actual
distribution (and effectively break the mass-inclination degeneracy).

\begin{figure}
\centerline{\plotone{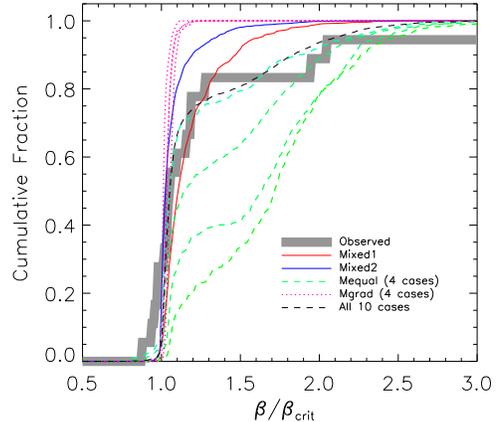}}
\caption{Cumulative distribution of $\bbc$ of the well-characterized
extra-solar multi-planet systems (in gray; see Table 1), as compared with our
scattering simulations. }
\label{fig:bdist}
\end{figure}

\section{Results}

We generated $\bbc$ distributions from our simulations following the procedure
described above.  First, we ``observed'' each system from 100 viewing angles,
thereby decreasing the inferred mass of each planet by a factor of $\sin
(I+i_j)$, where $i_j$ refers to each planet's inclination with respect to a
fiducial plane.  Second, we assumed the observed systems to be coplanar in
calculating $\bbc$ for each viewing angle using Eqn 1.  Finally, we included
the $\bbc$ calculated for each viewing angle by assuming the viewing angle $I$
to be isotropically distributed.  Figure~\ref{fig:bdist} compares the
cumulative $\bbc$ distributions for the observed two-planet systems with our
scattering simulations.  It is important to note that the ``true'' $\bbc$
distributions, calculated with knowledge of the simulated systems' real masses
and inclinations, are virtually identical to the curves from
Fig.~\ref{fig:bdist} (this issue is discussed further in \S 5).  Table 1 lists
the extra-solar systems in our analysis; we excluded systems with
controversial or poorly-characterized orbits and those that were likely
affected by tidal effects.  In a two planet system with an inner planet at
$\lesssim 0.1$ AU, tides will decrease the inner planet's eccentricity and
semimajor axis (Jackson \etal 2008), thereby increasing the separation between
the two planets and $\bbc$.

Four individual cases -- {\tt Mixed1, Mixed2, Meq:$\tt M_{Sat}$}, and {\tt
Meq:$\tt 30 \mearth$} -- each provide a match to the observed $\bbc$
distribution.  Kolmogorov-Smirnov (K-S) tests show that the probability $p$
that the $\bbc$ distributions from those four cases are drawn from the same
distribution as the observed sample are all 0.1 or larger (Table 2).  The
distribution calculated by an unweighted combination of all ten cases is also
a good match (each case was given equal weight, regardless of the number of
simulations).
\begin{deluxetable}{c|c|c|c|c}
\scriptsize
\tablewidth{0pc}
\tablecaption{Planetary systems included in $\bbc$ analysis\tablenotemark{1}}
\renewcommand{\arraystretch}{.6}
\tablehead{
\colhead{System} & 
\colhead{$a_1,a_2$} &
\colhead{$e_1,e_2$} &
\colhead{$M_1,M_2$} &
\colhead{$\bbc$} \\ 
\colhead{(pair)} &
\colhead{(AU)} &
\colhead{ } &
\colhead{($M_{Jup}$)} }
\startdata
HD 202206 b-c\tablenotemark{2} & 0.83,2.55 & 0.435,0.267 & 17.4,2.44 & 0.883\\
HD 82943 c-b\tablenotemark{2} & 0.746,1.19 & 0.359,0.219 & 2.01,1.75 & 0.946\\
HD 128311 b-c\tablenotemark{2} & 1.099,1.76 & 0.25,0.17 & 2.18,3.21 & 0.968\\
HD 73526 b-c\tablenotemark{2} & 0.66,1.05 & 0.19,0.14 & 2.9,2.5 & 0.982\\ HD
45364 b-c\tablenotemark{2} & 0.681,0.897 & 0.168,0.097 & 0.187,0.658 & 0.989\\
47 UMa b-c & 2.11,3.39 & 0.049,0.22 & 2.6,0.46 & 1.025\\ HD 155358 b-c &
0.628,1.224 & 0.112,0.176 & 0.89,0.504 & 1.043\\ HD 177830 c-b & 0.514,1.22 &
0.40,0.041 & 0.186,1.43 & 1.046\\ HD 60532 b-c\tablenotemark{2} & 0.77,1.58 &
0.278,0.038 & 3.15,7.46 & 1.054\\ HD 183263 b-c & 1.52,4.25 & 0.38,0.253 &
3.69,3.82 & 1.066\\ HD 108874 b-c\tablenotemark{2} & 1.051,2.68 & 0.07,0.25 &
1.36,1.018 & 1.10\\ HD 12661 b-c & 0.83,2.56 & 0.35,0.2 & 2.3,1.57 & 1.12\\ HD
11506 c-b & 0.639,2.43 & 0.42,0.22 & 0.82,3.44 & 1.17\\ HD 208487 b-c &
0.49,1.8 & 0.32,0.19 & 0.45,0.46 & 1.20\\ HD 169830 b-c & 0.81,3.60 &
0.31,0.33 & 2.88,4.04 & 1.28\\ HD 168443 b-c & 0.3,2.91 & 0.529,0.212 &
8.02,18.1 & 1.95\\ HD 38529 b-c & 0.129,3.68 & 0.29,0.36 & 0.78,12.7 & 2.06\\
HD 47186 b-c & 0.05,2.395 & 0.038,0.249 & 0.072,0.35 &  6.13
\enddata
\tablenotetext{1}{See
http://www.astro.washington.edu/users/rory/research/xsp/dynamics/ for an up to
date list of $\bbc$ values for the known extra-solar multiple planet systems.
Orbital values were retrieved from http://exoplanet.eu and
http://exoplanets.org.}
\tablenotetext{2}{These systems have been claimed to be in mean motion resonances.}
\end{deluxetable}

All of our sets of simulation produced a smaller fraction of systems at $\bbc
< 1$ than for the observed systems.  We therefore calculated K-S $p$ values by
confining the distributions to the ranges $\bbc \leq 1$ and $\bbc > 1$.  All
but one case with $p \geq 0.1$ also had $p \, (\bbc) > 0.1$ ({\tt Mixed2}; see
Table 2).  However, some cases provide good matches for $\bbc \leq 1)$ but not
for other regions, notably {\tt Meq:$\tt M_{Jup}$}, {\tt Mgrad:JSN}, and {\tt
Mgrad:NSJ}.  Systems with $\bbc < 1$ are unusual because they lie within the
formal Hill stability boundary but are stabilized by special orbital
configurations.  In fact, all five of the known exoplanet systems with $\bbc <
1$ are thought to lie in mean motion resonances (Table 1).  The scattered
systems with $\bbc < 1$ are stabilized by resonances or in many cases by
low-amplitude, aligned apsidal libration.  Four cases in our sample generated
resonant systems in at least 5\% of simulations (Raymond \etal 2008a) -- {\tt
Mixed2}, {\tt Mgrad:JSN}, {\tt Mgrad:NSJ}, and {\tt Mgrad:SJ3J} -- but only
two of these have $p \, (\bbc \leq 1) > 0.1$.  We attribute the lack of a
correlation between resonances and $\bbc < 1$ to the relative weakness of
these resonances.  Indeed, resonances caused by scattering tend to exhibit
relatively high-amplitude libration of only one resonant argument (Raymond
\etal 2008a); these resonances have typical $\bbc$ values of slightly more
than 1 (median $\bbc = 1.01-1.03$ for the different cases).  This contrasts
with resonances generated by convergent migration in gaseous disks, which tend
to exhibit low amplitude libration of more than one resonant argument
(Snellgrove \etal 2001; Lee \& Peale 2002).

Simulations with radial mass gradients ({\tt Mgrad}) overproduced systems very
close to the stability boundary, while cases with equal masses ({\tt Meq})
produced much larger $\bbc$ values (Fig.~\ref{fig:bdist}).  A similar effect
was seen in the eccentricity distributions: the {\tt Mgrad} cases yielded much
smaller eccentricities than the {\tt Meq} cases (Raymond \etal 2008a; see also
Ford \etal 2003).  The {\tt Mixed1} and {\tt Mixed2} cases fall between these
two regimes.  These trends can be explained by the number of close encounters
$n_{enc}$ that occur in the different cases before a planet is destroyed.  For
the {\tt Mgrad} cases $n_{enc}$ is typically between 30 and 80, and is larger
for less massive systems ({\tt JSN} and {\tt NSJ}).  For the {\tt Meq} cases
$n_{enc}$ is vastly larger, with median values between 100 ($\tt 3 M_{Jup}$)
and 2000 ($\tt 30 \mearth$).  The larger number of scattering events increases
the eccentricity of surviving planets and also causes the systems to spread
out farther.

\begin{deluxetable}{l|ccc}
\scriptsize
\tablewidth{0pc}
\tablecaption{$p$ values from K-S tests of observations vs. scattering simulations}
\renewcommand{\arraystretch}{.6}
\tablehead{
\colhead{Case} & 
\colhead{$p$} &
\colhead{$p\,(\bbc \leq 1)$} &
\colhead{$p\,(\bbc > 1)$}}
\startdata
\tt Mixed1 & 0.14 & $6.2 \times 10^{-3}$ & 0.53 \\
\tt Mixed2 & 0.13 & $1.2\times 10^{-4}$ & 0.07 \\
\tt Meq:3$\tt M_{Jup}$ & $1.8\times 10^{-6}$ & $2.1\times 10^{-5}$ & $4.6\times 10^{-4}$\\
\tt Meq:$\tt M_{Jup}$ & $9.3\times 10^{-4}$ & 0.10 & 0.02 \\
\tt Meq:$\tt M_{Sat}$ & 0.12 & 0.81 & 0.43\\
\tt Meq:30$\tt M_\oplus$ & 0.29 & 0.60 & 0.89\\
\tt Mgrad:JSN & $4.0\times 10^{-3}$ & 0.12 & $2.4\times 10^{-4}$\\
\tt Mgrad:NSJ & $1.5\times 10^{-4}$ & 0.10 & 0.02\\
\tt Mgrad:3JJS & $9.1\times 10^{-3}$ & $1.9\times 10^{-5}$ & $9.8\times 10^{-4}$\\
\tt Mgrad:SJ3J & $4.9\times 10^{-3}$ & $6.2 \times 10^{-4}$ & $2.8\times 10^{-4}$\\
All 10 cases & 0.14 & $5.1\times 10^{-3}$ & 0.58 
\enddata
\end{deluxetable}

\begin{figure}
\centerline{\plotone{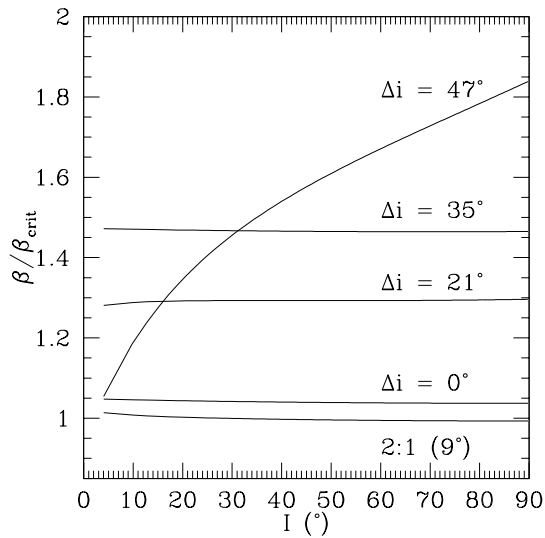}}
\caption{Inferred values for $\bbc$ as a function of observation
angle $I$ for several examples from the {\tt Mixed1} set, labeled by the
approximate mutual inclination $\Delta i$ between planets.  One resonant case
is labeled ``2:1'': it has $\Delta i = 9^\circ$. $I=90^\circ$ is edge-on and
$I=0^\circ$ is face-on. }
\label{fig:beta-i}
\end{figure}

In calculating ``observed'' $\bbc$ distributions from our simulations, we
assumed that the viewing angles $I$ were isotropically distributed.  Given
that known extra-solar planet systems are each observed at a fixed $I$, could
this have introduced a bias in our samples?  Figure~\ref{fig:beta-i} shows the
inferred value of $\bbc$ as a function of $I$ for five {\tt Mixed1} systems
with varying mutual inclinations $\Delta i$.  For $\Delta i \lesssim
35^\circ$, $\bbc$ varies only slightly with the viewing angle, but for large
$\Delta i$ $\bbc$ can change substantially with $I$.\footnote{We have found
that it is actually the angular momentum deficit (Laskar 1997) which controls
the magnitude of $\bbc$ variation with $I$.}  However, $\bbc$ varies by more
than 10\% [20\%] over the entire range of possible viewing angles for fewer
than 10\% [2\%] of cases.  For all systems, the edge-on $\bbc$ values agree
with the true $\bbc$ values (calculated with knowledge of the planets' true
masses and orbits) to better than 10\%.  There is a small bias: $\sim$80\% of
systems exhibit a shallow negative slope in $\bbc$ vs. $I$, suggesting that
the majority of inferred $\bbc$ values may be overestimated but only by
$\lesssim 1\%$.  Thus, although $I$ and $\Delta i$ are important to keep in
mind, they introduce a negligible error into our analysis.

\section{Discussion}

The planet-planet scattering model appears to be consistent with the $\bbc$
distribution of the observed exoplanet systems.  The distribution can be
reasonably reproduced by several of our sets of simulations, or even by an
unweighted combination of all ten sets.  We therefore cannot strongly
constrain the initial planetary mass distribution, although we can rule out
cases with very poor fits -- {\tt Meq:$\tt 3M_{Jup}$}, {\tt Mgrad:3JJS} and
{\tt Mgrad:SJ3J} in particular -- as the major contributors to the
distribution (see Table 2).  We consider the {\tt Mixed1} set to be the most
realistic because it is drawn from the observed mass distribution (Butler
\etal 2006), and it provides a good match to the observed eccentricity
distribution (Raymond \etal 2008a).  In the coming years, we expect many more
systems to be discovered with $\bbc \approx 1-1.5$.

The pileup of scattered systems just beyond the stability boundary implies
that planetary systems are ``packed'', meaning that large spaces in between
planets should be rare.\footnote{It is important to note that the spacing for
planets with $\bbc \approx 1$ can be large.  Among just the {\tt Mixed1}
simulations with $1 \leq \bbc \leq1.1$ the difference in semimajor axis for
adjacent planets ranges from $<$2 AU to $>$15 AU.}  This provides a
theoretical foundation for the ``Packed Planetary Systems'' hypothesis, which
asserts that if a stable zone exists between two known planets, then that zone
is likely to contain a planet (Barnes \& Raymond 2004; Raymond \& Barnes 2005;
Raymond \etal 2006).  Given the small $\bbc$ values of scattered systems,
there is simply no room to insert another planet between the two known planets
without causing the system to be unstable.

HD 74156 is an example of a packed planetary system.  Prior to 2008, two
planets were known in the system, at 0.28 and 3.4 AU (Naef \etal 2004), with $\bbc = 1.987$.
Raymond \& Barnes (2005) mapped out a narrow stable zone between the two
planets, from 0.9-1.4 AU.  The planet HD 74156 d was discovered three years
later by Bean \etal (2008) at 1.01 AU (see also Barnes \etal 2008) at the peak
of the stable zone.  We therefore expect additional planets to exist in
systems with $\bbc > 1.5-2$, notably HD 38529 (Raymond \& Barnes 2005) and HD
47186 (Kopparapu \etal 2009).  The probable location of additional planets can
be determined using test planets to map out dynamically stable regions between
known planets (e.g., Menou \& Tabachnik 2003; Rivera \& Haghighipour 2007;
Raymond \etal 2008b).

The $\bbc$ distribution of the observed extra-solar planetary systems may
contain information about different dynamical regimes.  The region of $\bbc
\leq 1$ is populated entirely by resonant systems and may provide evidence of
planetary system compression, presumably via convergent migration in gaseous
protoplanetary disks (Snellgrove \etal 2001; Lee \& Peale 2002).  The region
of $1 \leq \bbc \leq 1.5-2$ is consistent with the scattering regime.
Widely-separated systems with $\bbc > 1.5-2$ may have been drawn apart by
interactions with planetesimal or gaseous disks (e.g., Gomes \etal 2004;
Moeckel \etal 2008).  However, this seems unlikely given that the known
planets lie relatively close to their stars and that disk effects should be
far more pronounced at large distances.

Given that our simulations started with only three planets, we could not
calculate $\bbc$ values in perturbed two-planet systems.  For example, an
interesting comparison with observations would be to measure $\bbc$ for the
two easiest-to-detect planets in scattered three planet systems.  This would
address the question of whether to search for additional planets in between or
interior/exterior to the known planets in two planet systems with large
$\bbc$.

\vskip .2in
We thank Google for access to their machines.  S.N.R. and R.B. acknowledge
funding from NASA Astrobiology Institutes's Virtual Planetary Laboratory lead
team, supported by NASA under Cooperative Agreement No. NNH05ZDA001C.


\end{document}